\documentclass{ws-ijbc}
\usepackage{graphicx}
\usepackage{multirow}
\usepackage{dcolumn}
\usepackage{amsmath}
\usepackage{comment}

\graphicspath{{figures/}}

\begin{document}

\markboth{J. Roulet \& G. B. Mindlin}{A Diagrammatic Representation of Phase Portraits and Bifurcation Diagrams of Two-Dimensional Dynamical Systems}

\title{A DIAGRAMMATIC REPRESENTATION OF PHASE PORTRAITS AND BIFURCATION DIAGRAMS OF TWO-DIMENSIONAL DYNAMICAL SYSTEMS}

\author{JAVIER ROULET}
\address{Physics Department, Princeton University\\
Princeton, NJ 08544, USA\\
jroulet@princeton.edu}

\author{GABRIEL B. MINDLIN}
\address{Departamento de F\'isica, Facultad de Ciencias Exactas y Naturales, Universidad de Buenos Aires,\\
Buenos Aires, C1053ABJ, Argentina\\
and IFIBA, CONICET\\
Ciudad Universitaria, 1428 Buenos Aires, Argentina\\
gabo@df.uba.ar}

\maketitle

\begin{abstract}
We treat the problem of characterizing in a systematic  way the qualitative features of two-dimensional dynamical systems. To that end, we construct a representation of the topological features of phase portraits by means of diagrams that discard their quantitative information. All codimension 1 bifurcations are naturally embodied in the possible ways of transitioning smoothly between diagrams. We introduce a representation of bifurcation curves in parameter space that guides the proposition of bifurcation diagrams compatible with partial information about the system.
\end{abstract}


\section{Introduction}

By understanding the dynamics displayed by a nonlinear system we typically refer to the capacity to list all the qualitatively different phase space portraits that the system can display for different values of its control parameters. Even when the equations ruling the system are known, it is often a very hard problem. Yet, there are algorithmic ways to proceed. One computes some key invariant sets, analyzes their stability, finds the normal forms that allow mapping the problem onto a (hopefully) studied one close to a bifurcation\dots until consistent bifurcation diagrams are sketched for all the parameters of interest \cite{wiggins2003, Guckenheimer1983}. Eventually, the educated intuition of a dynamicist allows filling a gap, so that every single change in the phase portrait, as the parameters are changed, can be explained by either a local or global bifurcation.  The program becomes much more difficult when the equations are not known, for example, if one explores the problem experimentally \cite{Green1990,DAngelo1992,Valling2007,Ondarcuhu1993,Ondarcuhu1994,Mindlin1994,Berry1996}. In that case, one starts with some sets of attractors, obtained for different parameter values, which a priori are not ``close'' in any way. Actually, a similar situation is faced when a system (whose equations are known) is explored numerically. Is it possible to algorithmically list and classify the dynamical possibilities compatible with sparse information of this sort?

In this work we explore this question for planar systems. These systems are near and dear to the hearts of dynamicists, since two is the minimal dimensionality in which we can embed nontrivial, recurrent dynamics. Moreover, it is typical to study the different behaviors that these bidimensional models can display when two parameters are varied, since this allows us to consider cases in which the linear part of the vector fields are doubly degenerate. Yet, even these modest models can present a significant puzzle for a natural scientist designing the set of experiments (or numerical simulations) necessary to unveil the structure of his/her problem's bifurcation diagram. The tools we present here provide an algorithmic means for generating and classifying all phase portraits compatible with a given, limited information about a dynamical system. This could be for instance the knowledge from experiments or simulations about what the attractors of the system are.

The work is organized  as follows. In Section~\ref{sec:diagrams} we introduce a way of representing phase portraits by using diagrams that capture important qualitative information of the system's dynamics. Specifically, they encode what the limit sets of the system are, their stability and their distribution in phase space.

Section~\ref{sec:bifurcations} discusses how smooth modifications of these diagrams give rise naturally to bifurcations, in which phase portraits change qualitatively. All codimension 1 bifurcations are obtained in this way. We introduce a representation of bifurcation curves by means of ``dressed'' lines, that encode the direction in which new limit sets are created and their type and stability. The possible ways of connecting different bifurcation curves in a higher codimension bifurcation can be constrained by simple rules concerning their dressings.

In Section~\ref{sec:examples} we show, as an instructive example, how the theoretical framework developed here can be applied to the Wilson-Cowan oscillator. We present our conclusions in Section~\ref{sec:conclusions}.

\section{Diagram Representation of Phase Portraits}
\label{sec:diagrams}

We now describe a representation of phase portraits by diagrams that discard all the quantitative information that portraits convey (i.e., the specific trajectories in phase space), while preserving the qualitative features. Unlike the phase portrait, the resulting diagram is robust under changes in the system's parameters as long as these do not reach a bifurcation, in which the qualitative features of the system change. 

We will restrict ourselves to two-dimensional, structurally stable dynamical systems. Further, we will assume that the region of interest of phase space can be enclosed by a closed transversal curve (i.e., a closed curve along which the velocity vector is neither tangent nor zero, so it always traverses the curve from one of the sides to the other). This is, we assume that the flow traverses the whole boundary of the region of interest either inward or outward. We note that the structural stability hypothesis is usually generic except for systems that have some kind of symmetry or conserved quantity, to which this method does not apply in a straightforward manner.

\subsection{Construction of the Diagrams}
\label{ssec:construction}

The diagrams encode information about the limit sets of the system, which in two dimensions can only be stable or unstable nodes (or foci, which are topologically equivalent), stable or unstable limit cycles, or saddle points. We will represent those with the symbols of Table~1. 

\begin{table}[h]
	\tbl{Diagram representation of the limit sets possible in two-dimensional dynamical systems.}
	{\begin{tabular}{llD{.}{.}{3.2}D{.}{.}{3.2}}\\
		\toprule
		Limit set & Representation & \multicolumn{1}{c}{Index} & \multicolumn{1}{c}{Repulsion} \\
		\hline
		\begin{tabular}{@{}l@{}}Stable\\node\end{tabular} & 
		\begin{minipage}{.33\linewidth}
			\includegraphics[height=1.3cm]{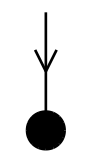}
		\end{minipage}&
		1 & -1\\
		\begin{tabular}{@{}l@{}}Unstable\\node\end{tabular} & 
		\begin{minipage}{.33\linewidth}
			\includegraphics[height=1.3cm]{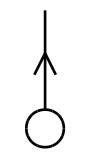}
		\end{minipage}&
		1 & 1\\[15pt]\hline
		\begin{tabular}{@{}l@{}}Stable\\cycle\end{tabular} & 
		\begin{minipage}{.33\linewidth}
			\includegraphics[height=1.3cm]{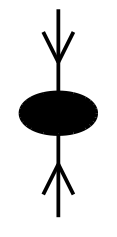}
		\end{minipage}&
		0 & -2\\
		\begin{tabular}{@{}l@{}}Unstable\\cycle\end{tabular} & 
		\begin{minipage}{.33\linewidth}
			\includegraphics[height=1.3cm]{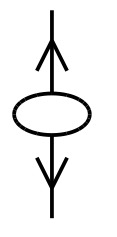}
		\end{minipage}&
		0 & 2\\[15pt]\hline
		\multirow{4}{*}{Saddle} & 
		\begin{minipage}{.33\linewidth}
			\includegraphics[height=1.3cm]{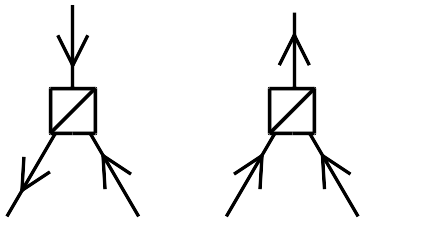}
		\end{minipage}&
		-1 & -1\\
		& 
		\begin{minipage}{.33\linewidth}
			\includegraphics[height=1.3cm]{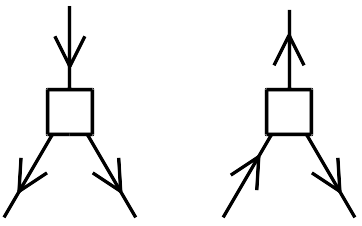}
		\end{minipage}&
		-1 & 1\\[15pt]\hline
		2 saddles & 
		\begin{minipage}{.33\linewidth}
			\includegraphics[height=1.3cm]{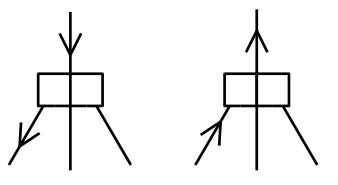}
		\end{minipage} & \multicolumn{2}{c}{Additive} \\
		\qquad\vdots & \\
		$n$ saddles & 
		\begin{minipage}{.33\linewidth}
			\includegraphics[height=1.3cm]{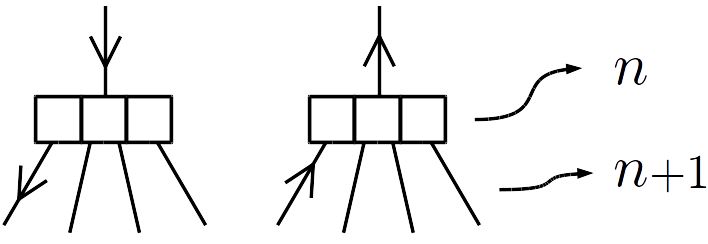}
		\end{minipage} & \multicolumn{2}{c}{Additive} \\
		\botrule
	\end{tabular}}
\end{table}

In Table~1 we have introduced two quantities associated to limit sets: the index and the repulsion. These can be computed easily from the diagrams: the index is the number of arrows above its symbol minus the arrows below, and the repulsion is the number of outgoing arrows minus the incoming ones. Their interpretation will be given in \S\ref{ssec:index_repulsion}.

The shape of a symbol encodes the type of limit set it represents, and the filling its repulsion. For nodes and cycles, the repulsion is directly related to their (in)stability.

Notice that there are various ways in which the three arrows connecting the square saddle point symbol can be oriented. It is forbidden to have all the arrows departing from the square, or all converging, but the other possibilities are allowed. These represent different ways in which the saddle's invariant manifolds can be connected to other limit sets, as explained later (\S\ref{ssec:diagram2portrait}), and determine its repulsion: $-1$ for symbols with two incoming arrows and one outgoing, and $1$ for the converse.

If a system has several saddle points, we shall represent them with a compound symbol as shown in the last two entries of Table~1 (unless they are separated by a limit cycle so that one is inside the limit cycle and the other outside, in which case we treat them individually). In those symbols, the undirected lines should be replaced with incoming or outgoing arrows, and one of the squares should be crossed for each extra incoming arrow (see Fig.~\ref{fig:examples}c for an example). As before, it is forbidden to have all the $n+2$ arrows outgoing or all incoming. 

Every well-formed diagram consists of any number of the symbols in Table~1, connected by arrows so that there is only one arrow unmatched, at the top of the diagram. Each connecting arrow should have a definite direction (upward or downward), so two symbols can only be connected if the directions of their arrows match. For example, some well-formed diagrams are shown in Fig.~\ref{fig:examples}.

Suppose we know from experiments or simulations what the stable limit sets of a two-dimensional dynamic system are. Then it is possible to obtain all the phase portraits compatible with these stable solutions by adding unstable sets (i.e. unstable nodes, unstable cycles or saddle points) in a way such that the resulting diagrams are well-formed. Each diagram can then be interpreted as a specific class of phase portraits.

\subsection{Interpretation of the Diagrams as Phase Portraits} \label{ssec:diagram2portrait}

\begin{figure}
\centering
\includegraphics[width=.5\linewidth]{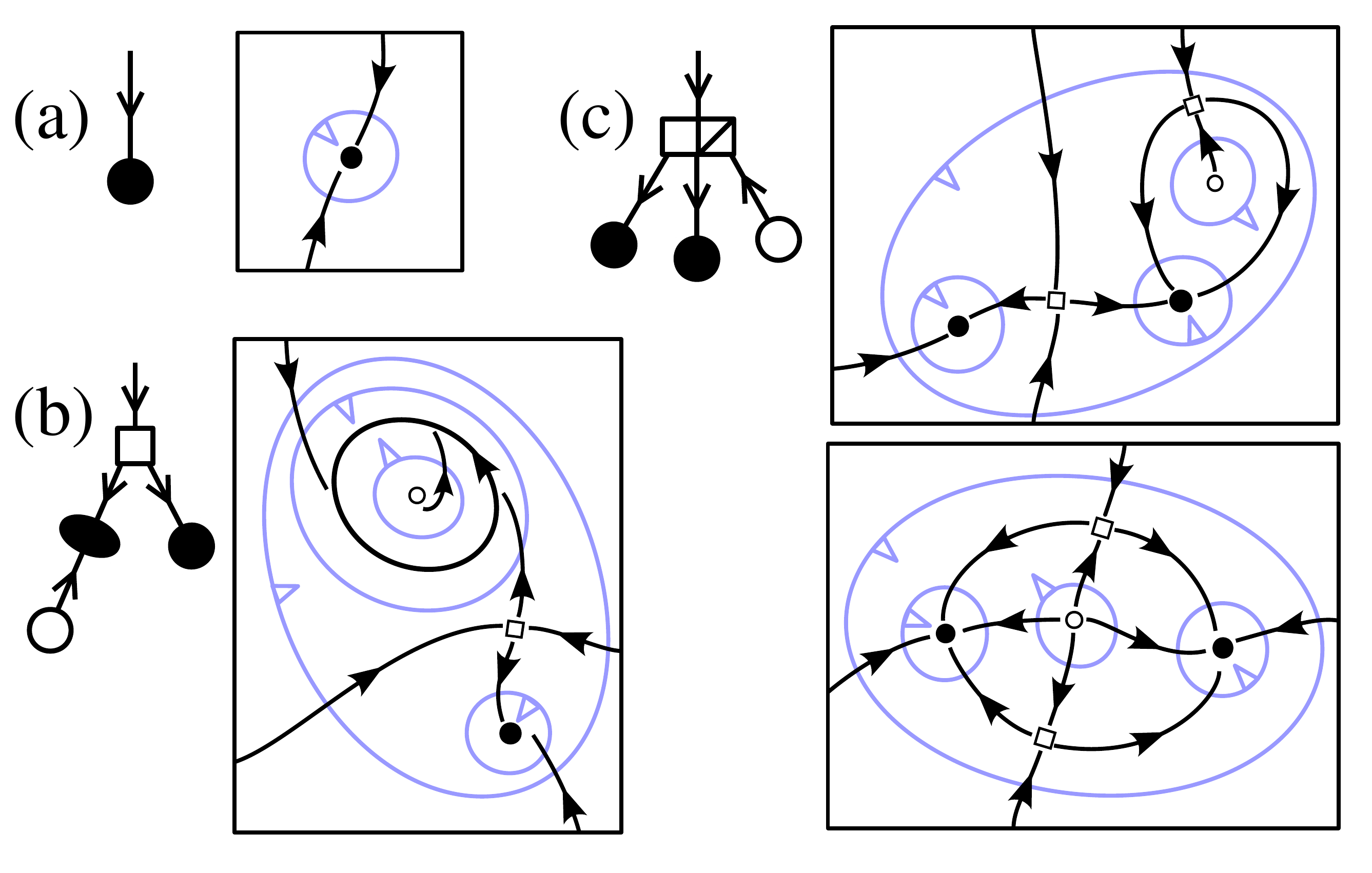}
\caption{Some examples of well-formed diagrams and their interpretation as phase portraits. In the portraits, filled dots represent stable nodes, empty dots unstable nodes, squares saddle points, and closed curves limit cycles. Closed transversals are shown in blue, with an arrowhead indicating the inward or outward direction of the flow. Each closed transversal is associated to an arrow in the diagram. Diagram (c) admits multiple topologically different phase portraits, that would be separated by heteroclinic bifurcations in parameter space.}
\label{fig:examples}
\end{figure}

The interpretation of the diagrams is as follows: each arrow in the diagram represents a family of closed transversal curves in the phase plane, that enclose all the limit sets in the branch of the diagram below that arrow (see insets in Fig.~\ref{fig:examples}). An arrow pointing downwards means that the flow enters the region delimited by the associated closed transversal, and an arrow pointing upwards means that it exits it. Given a diagram, we draw a closed curve in phase space for each arrow, keeping track of its direction by adding an inward or outward arrowhead to the curve. We draw each curve inside the one associated to the arrow immediately above in the diagram.

Notice that every symbol in Table~1 has one arrow on top and a number of arrows at the bottom. Accordingly, the transversal curves in phase space will delimit regions with an outer boundary and a number of holes inside. In each of these regions we draw the limit sets associated with the corresponding symbol in the diagram. Regions with a limit cycle will have a single hole, and the limit cycle should be drawn around it (see Fig.~\ref{fig:examples}b).

\begin{figure}
    \centering
    \includegraphics[width=.5\linewidth]{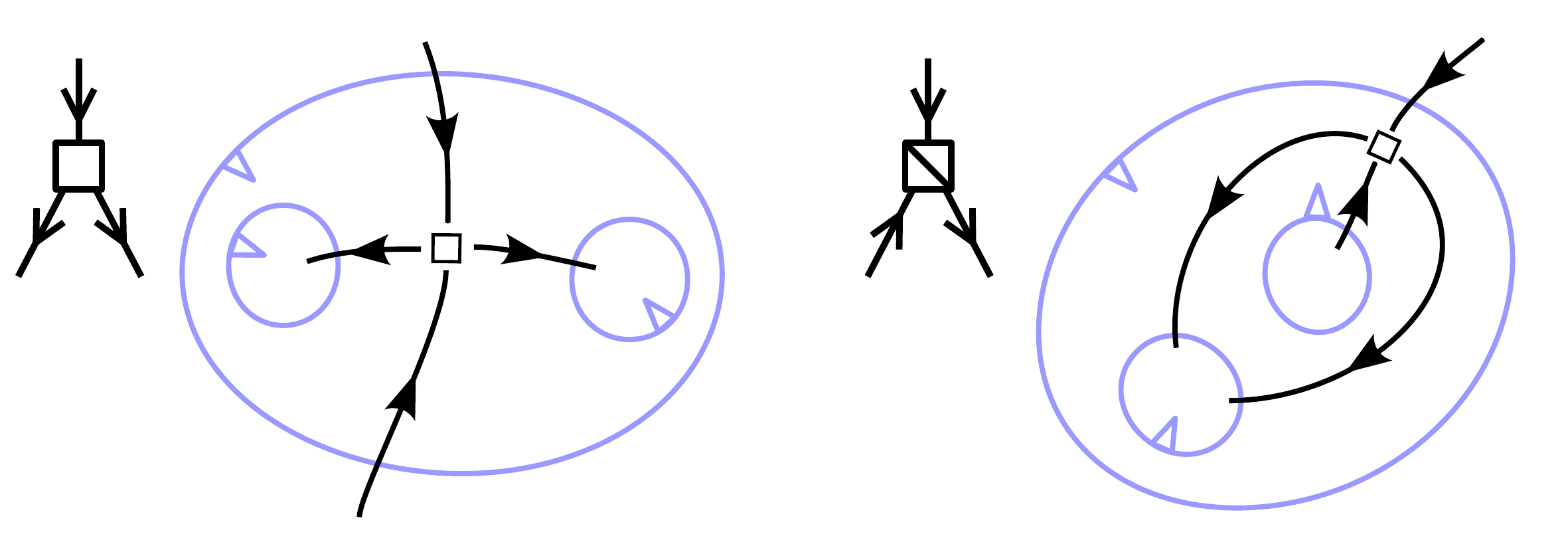}
    \caption{Two different ways in which the invariant manifolds of the saddle point can be connected to other closed transversals. The remaining two can be obtained from these by inverting the direction of all the arrows and trajectories.}
    \label{fig:saddle_types}
\end{figure}

To complete the phase portrait, trajectories that cross all the transversals in the direction they define should be sketched. Of these, the ones along the saddles' invariant manifolds are the most interesting. Two begin in each saddle and two end, constituting its unstable and stable manifold respectively. Each pair approaches or leaves the saddle in opposite directions, so the incoming and outgoing trajectories alternate around it. For a single saddle point, there is a unique qualitative (topologically equivalent) way of connecting its manifolds to the transversals bounding its region, that depends on its repulsion as shown in Fig.~\ref{fig:saddle_types}. This follows since these trajectories must each traverse one of the bounding transversals in the direction it induces, without intersecting the other trajectories. For several saddle points, the ways of connecting their invariant manifolds to the transversals are no longer unique. However, all of them are equivalent up to topological equivalence and heteroclinic bifurcations. An example is shown in Fig.~\ref{fig:examples}c.

In the case of multiple saddle points, we do not attempt to assign repulsions to them individually, because it cannot always be done without ambiguity (see \S\ref{ssec:index_repulsion}). A repulsion can be assigned, however, to the whole set of saddles. By doing this, our treatment abandons a complete description of the connectivities of the invariant manifolds of the saddle points, and thereby ignores heteroclinic bifurcations. We favor this choice over complicating the notation to account for the detailed connectivity of multiple saddles, because in any case that information is not usually accessible to experiments.
 
The only other topological ambiguity left when reconstructing the phase portrait is the sense in which the limit cycles circulate. If this is relevant for a certain application, it can be easily incorporated to the method by using different symbols for clockwise- or counterclockwise-circulating cycles. For simplicity, we will ignore this distinction here.

Thus, a specific diagram represents a class of phase portraits, all of which equivalent up to topological equivalence and, if several saddles are present, heteroclinic bifurcations. In the following section we will show the converse statement: that every phase portrait satisfying the hypotheses made at the beginning of this section is represented by a unique diagram.

\subsection{Obtaining the Diagram from the Phase Portrait}

Given a phase portrait, the diagram that represents it can be constructed following algorithmic steps. We need to identify the closed transversals on the phase portrait and associate an arrow in the diagram to each of them. The arrows then define the limit set symbols according to Table~1.

For every stable or unstable node, a sufficiently small transversal can always be found that encircles it, we draw one of these with the appropriate orientation for each node present (inward flow for stable, outward for unstable). Similarly, for every stable or unstable limit cycle a pair of closed transversals can always be found sufficiently close to the cycle, one on the inside and the other on the outside. Finally, by hypothesis the entire region of interest can be enclosed with a closed transversal, which we also draw. All of these transversals can -- and should -- be chosen so that they do not intersect. As a result, the closed transversals will define regions with an outer boundary and a number of holes. It could be the case that a region has exactly one hole and the two bounding transversals are oriented both inward or both outward. Such regions cannot have any limit set inside, we eliminate them by discarding either of the two bounding transversals. 

After this process, the phase space will be divided by the transversals in regions enclosing a single node, a single cycle or a nonzero number of saddle points. That is, matching one of the entries of Table~1. The full diagram can now be constructed following the hierarchy induced by the distribution of these regions. The symbol of a region lying inside a hole of another should be located below the outer region's symbol, and connected with an arrow in the direction induced by the orientation of the transversal separating them.

\bigskip

Thus, we have an algorithmic way of identifying any phase portrait with a specific diagram, and any diagram with a specific class of phase portraits. The diagrams may be used to consistently generate and classify phase portraits in a highly qualitative approach, and restrictions about the system's limit sets can be naturally applied to them.

\subsection{Index and Repulsion} \label{ssec:index_repulsion}

We can now give an interpretation to the index and repulsion introduced in Table~1. In planar dynamical systems, the index of a closed curve is defined as the amount of counterclockwise revolutions that the vector field does as one travels counterclockwise once around the curve. In particular, the index of a closed transversal is always $1$. To compute the index of a limit set, we extend that definition in the following way: we first choose a region of phase space that contains the limit set we are interested in and no other, and that is bounded by closed curves. The index of the limit set is the sum of the indexes of all the bounding curves, with the following proviso: the direction for moving around the bounding curves should be with the region to the left, i.e. counterclockwise for ``outer'' boundaries but clockwise for inner, ``hole'' boundaries, which gives the opposite sign. For example, the simplest region containing a limit cycle and no other limit sets has a ring shape with an outer boundary and an inner hole, which can be chosen transversal to the flow. The outer boundary should be traveled counterclockwise, giving an index of $1$, and the hole clockwise yielding index $-1$. The index of limit cycles is thus $0$. Since the index of a curve is invariant under continuous deformations of it that do not traverse a fixed point, the choice of a boundary transversal to the flow is not necessary. In general, the index of a region bounded by transversal curves is the number of outer boundaries minus the number of holes, which justifies the prescription of obtaining it from the diagram by subtracting the number of arrows below the symbol to the arrows above.

We define the repulsion only for regions bounded by closed transversals, and it is directly the number of transversals through which the flow leaves the region minus the number through which it enters it. The repulsion of a limit set is the repulsion of a region bounded by closed transversals that contains it and no other limit sets, where we are using ``limit set'' to actually refer to any of the entries of Table~1. The subtlety is that if a system has multiple saddle points, a repulsion cannot always be assigned unambiguously to individual saddle points. For instance, in the lower inset of Fig.~\ref{fig:examples}c, we could enclose both stable nodes and either of the two saddles with a new closed transversal, which naively would give that saddle a repulsion of $+1$, and $-1$ to the other, depending on our arbitrary choice of the transversal. The pair of saddles taken as a whole, however, has repulsion $0$ (in this example) regardless of that choice. We can then simply treat both saddles as a single ``limit set'' with repulsion 0 and index $-2$, motivating the convention for the symbols of Table~1.

Both quantities are additive: the index of the union of two disjoint regions is the sum of their indexes, and similarly for the repulsion. The interesting case is when two symbols adjacent in the diagram are considered together: then the arrow that connects them is not counted when computing these quantities. However, the arrow must have been above one of the symbols and below the other, so the contribution to the index of one of the limit sets cancels the contribution to the other, yielding the same total index. Similarly, the connecting arrow is necessarily outgoing for one of the symbols and incoming for the other, so it does not contribute to the total repulsion either.

An important observation is that for the class of dynamical systems we are considering, i.e. whose relevant phase diagram can be bounded by a single closed transversal, the total index is $1$ and the total repulsion is $-1$ if the flow enters the transversal, or $1$ if it exits it. These quantities can also be obtained directly from the limit sets of the system, which has important consequences. First, from Table~1 we see that the total index is the number of nodes minus the number of saddles, so we conclude that the nodes must always exceed the saddles by one. In particular, these systems have an odd number of fixed points, which can be used as a criterion to assess whether all of them have been found in a search. Second, the sum of the repulsions of the limit sets is also given. In particular, the repulsion of a saddle point might be assigned a priori given the rest of the limit sets, which can be used to constrain the ways in which its invariant manifolds are distributed in phase space (Fig.~\ref{fig:saddle_types}). These constraints are naturally encoded by the diagrams, since well formed diagrams have only one external arrow on top, yielding a total index of $1$ and a total repulsion given by its orientation. All the other arrows are connected at both ends, giving no contribution.

\section{Bifurcations}\label{sec:bifurcations}

\subsection{Continuous Transitions between Diagrams} \label{ssec:transitions}

Varying the system's parameters continuously may change its behavior qualitatively, a process known as bifurcation. Then, in crossing a bifurcation the diagram describing the system should change, and do so somehow ``continuously''. Interestingly, the diagram formalism allows a natural interpretation of all bifurcations as the continuous ways to change a diagram into another. 

By changing a diagram continuously we mean either shrinking the length of a connecting arrow until it disappears, or alternatively creating a new zero-length arrow and enlarging it. Since the arrows imply the limit set symbols according to Table~1, these must be updated as the arrow configuration changes.

Recall that each arrow represents a family of closed transversals that separate the limit sets associated to the symbols in its ends. An arrow length approaching zero is interpreted as the two involved limit sets approaching each other in phase space, so that a transversal curve should be finely tuned to separate them. In the bifurcation, the limit sets collide and no separating curves can be found any more.

Depending on the type of limit sets that are connected to the arrow involved, different kinds of bifurcation can occur. Table~2 shows all the possible connections between symbols that can be made using the entries of Table~1, and the transition to a different diagram that takes place when the involved arrow shrinks to zero-length. Each of these transitions has an interpretation as a bifurcation. The (partial) diagrams of Table~2 display only the limit sets involved in the bifurcation. They must be completed with other limit sets, which would not participate in the bifurcation, in order to represent a full phase portrait. For instance, at the bottom of the first diagram there is a downward arrow left unmatched, that could be completed by adding a stable node at both sides of the transition.

\begin{table}[ht!]
    \centering
    \tbl{All possible continuous transitions between different diagrams, each involves an arrow whose length approaches zero and represents a codimension 1 bifurcation. Conversely, all codimension 1 bifurcations in two-dimensional systems that change the limit sets can be expressed as a continuous transition between diagrams.}
    {\begin{tabular}{lc}
    	\toprule
    	Bifurcation & Transition between diagrams \\\hline
    	Saddle-Node & 
    	\begin{minipage}{.3\linewidth}\centering % eg .5 in two-column
    		\includegraphics[width=\linewidth]{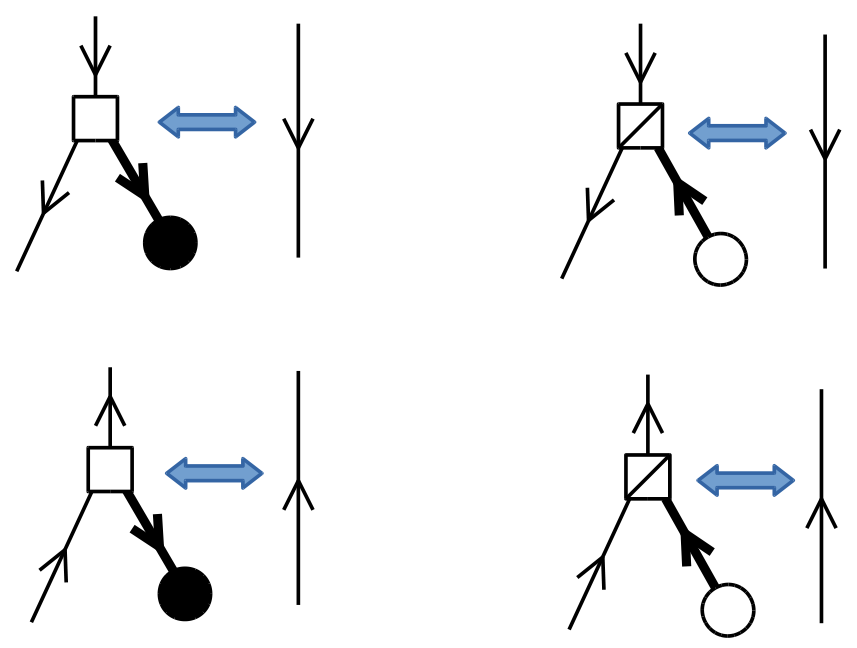}
    	\end{minipage}\\[15pt]\hline
    	Hopf &
    	\begin{minipage}{.3\linewidth}\centering
    		\includegraphics[width=\linewidth]{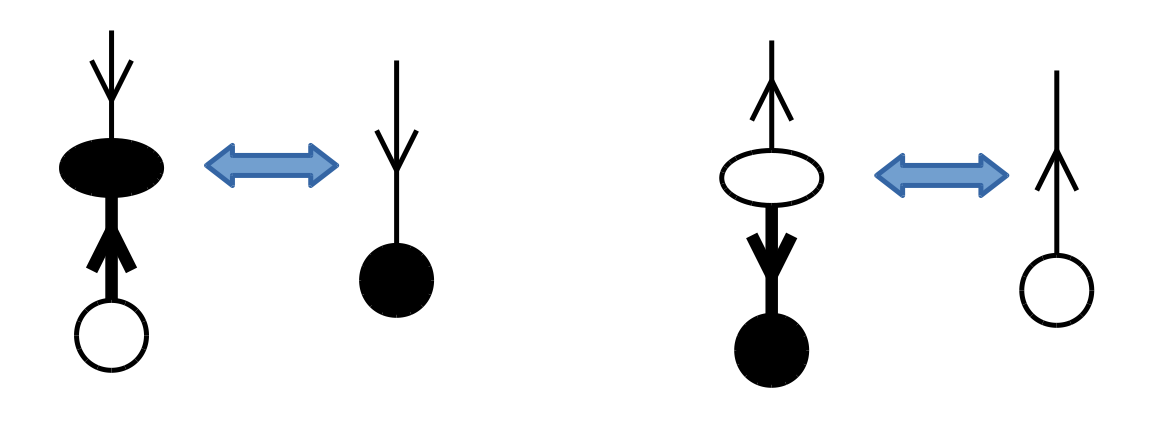}
    	\end{minipage}\\[15pt]\hline
    	\begin{tabular}{@{}l@{}}Saddle-Node on\\Invariant Cycle\end{tabular} & 
    	\begin{minipage}{.3\linewidth}\centering
    		\includegraphics[width=\linewidth]{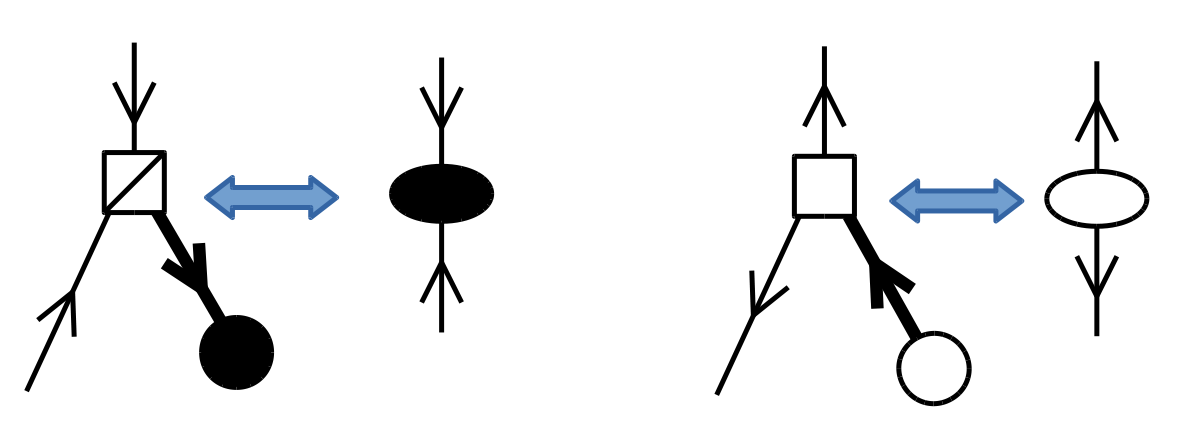}
    	\end{minipage}\\[15pt]\hline
    	Homoclinic &
    	\begin{minipage}{.3\linewidth}\centering
    		\includegraphics[width=\linewidth]{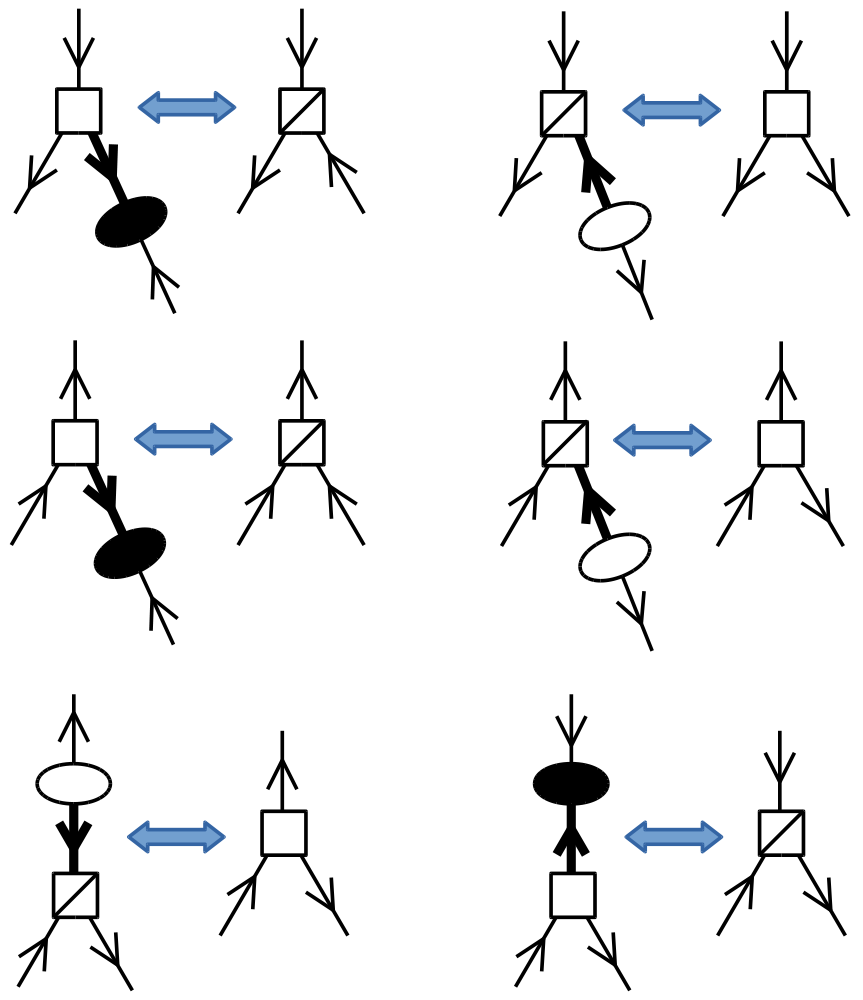}
    	\end{minipage}\\[15pt]\hline
    	\begin{tabular}{@{}l@{}}Saddle-Node of\\Limit Cycles\end{tabular} & 
    	\begin{minipage}{.3\linewidth}\centering
    		\includegraphics[width=\linewidth]{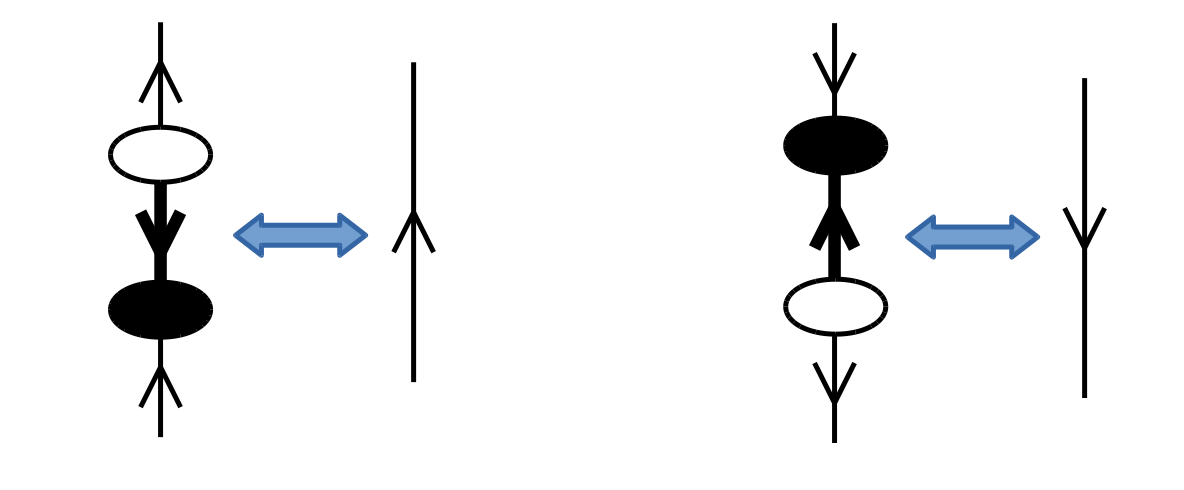}
    	\end{minipage}\\[15pt]
    	\botrule
    \end{tabular}}
\end{table}

Notice that the amount and orientation of all the arrows external to each partial diagram remains unchanged in the bifurcations. This yields conservation laws for the index and repulsion of the system, which must be constant for all parameter values as long as the hypotheses made at the beginning of Section~\ref{sec:diagrams} still hold. Some remarkable consequences follow from this: we can predict a priori from the diagram whether a collision between a saddle and a node would lead to a regular saddle-node bifurcation (if they have opposite repulsions) or to a saddle-node on invariant cycle (if they have the same repulsion). Similarly, from the repulsion of a saddle point we can predict the stability of a limit cycle born from it at a homoclinic bifurcation (and, conversely, whether a limit cycle of given stability can annihilate against it). Moreover, the specific distribution of arrows around the saddle point symbol determines the topologically allowed ways in which the cycle can be created, in particular, whether ``big homoclinic loops'' can occur or not.

\subsection{A Representation of Bifurcations in Parameter Space}

We will now introduce a representation of codimension 1 bifurcations in parameter space, that explicits which limit sets are created or destroyed in crossing the bifurcation. This helps proposing plausible bifurcation diagrams from partial knowledge of the behavior of the system, in a way analogous to the diagrams introduced in Section~\ref{sec:diagrams}, that allowed to systematically construct plausible phase portraits. We will focus on two-dimensional parameter spaces, in which codimension 1 bifurcations are curves.

We can keep track of the creation and annihilation of limit sets at a bifurcation by ``dressing'' its curve with symbols at its sides. Table~3 shows our convention, in which triangles represent nodes, loops represent cycles and squares represent saddle points, and the filling indicates their repulsion. For example, a filled triangle on the right side of a curve would indicate that a stable node is created when crossing the bifurcation from left to right, or destroyed if crossed from right to left. The symbols associated to each curve can be obtained from Table~2 by identifying the limit sets that intervene in the bifurcation. They add to the same index and repulsion at each side of the curve, so that these conserved quantities do not change when the system undergoes the bifurcation. 

\begin{table}
\centering
\tbl{Representation of codimension 1 bifurcations in a two-dimensional parameter space. The bifurcations are represented by dressed curves, the dressing symbols indicate what limit sets are created or destroyed in the bifurcation.}
{\begin{tabular}{lc}
	\toprule
	Bifurcation & Representation in parameter space \\\hline
	Saddle-Node & 
	\begin{minipage}{.25\linewidth}\centering % Change to eg .5 for two-column
		\includegraphics[width=\linewidth]{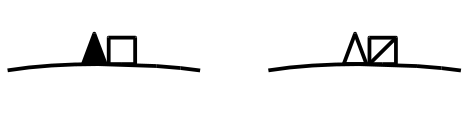}
	\end{minipage}\\
	Hopf & 
	\begin{minipage}{.25\linewidth}\centering
		\includegraphics[width=\linewidth]{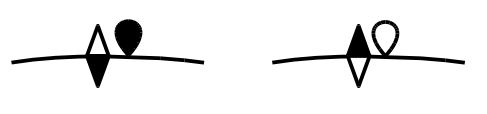}
	\end{minipage}\\
	\begin{tabular}{@{}l@{}}Saddle-Node on\\Invariant Cycle\end{tabular} & 
	\begin{minipage}{.25\linewidth}\centering
		\includegraphics[width=\linewidth]{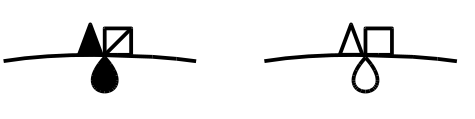}
	\end{minipage}\\
	Homoclinic & 
	\begin{minipage}{.25\linewidth}\centering
		\includegraphics[width=\linewidth]{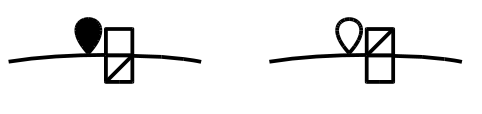}
	\end{minipage}\\
	\begin{tabular}{@{}l@{}}Saddle-Node of\\Limit Cycles\end{tabular} & 
	\begin{minipage}{.25\linewidth}\centering
		\includegraphics[width=.5\linewidth]{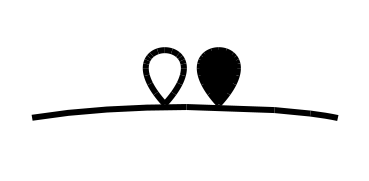}
	\end{minipage}\\
	\botrule
\end{tabular}}
\end{table}

\begin{figure}[h]
	\centering
	\includegraphics[width=.45\linewidth]{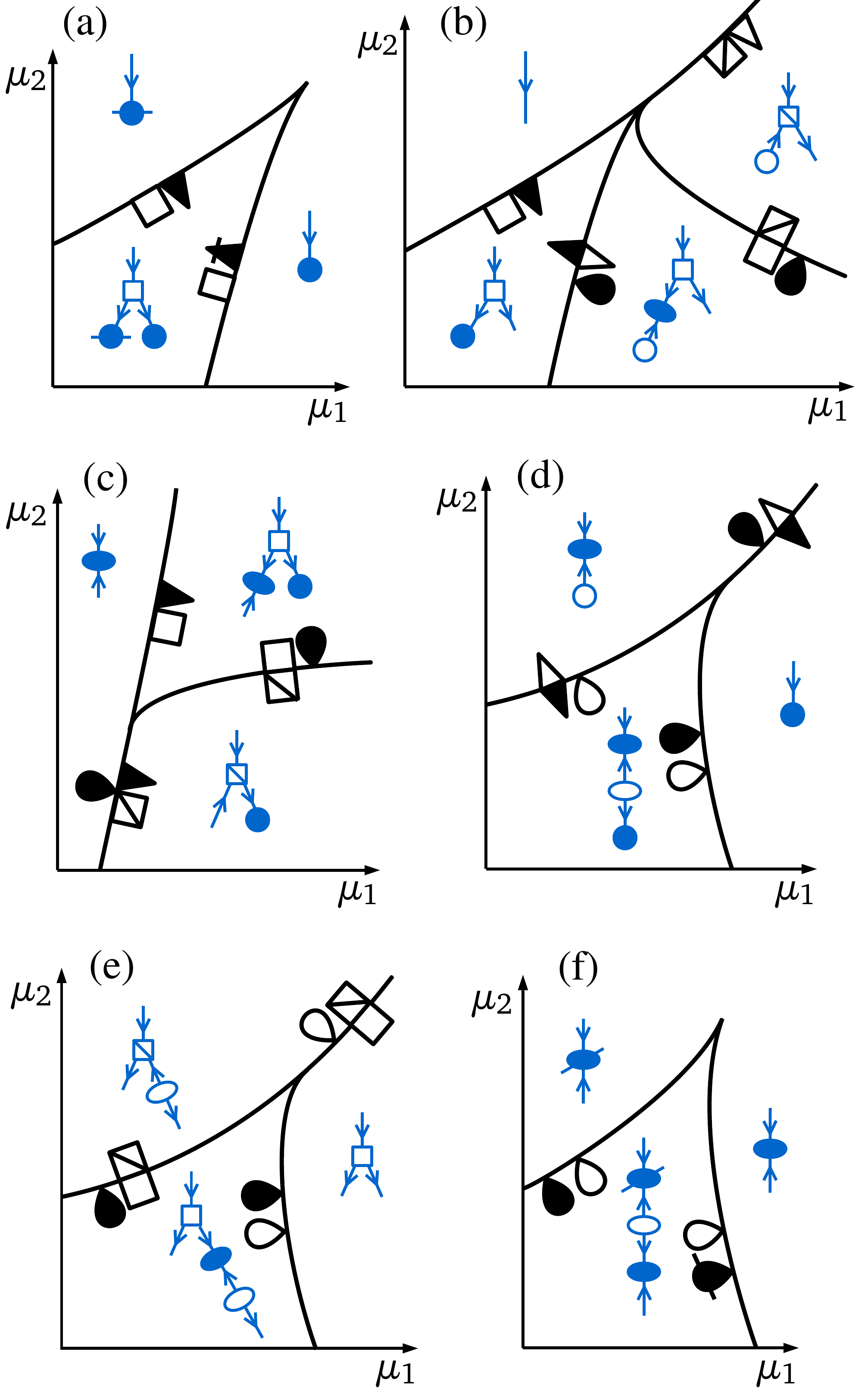}
	\caption{Representation of codimension 2 bifurcations in a two-parameter space $(\mu_1, \mu_2)$, at the meeting of codimension 1 bifurcation curves. These correspond to (a) cusp, (b) Bogdanov-Takens, (c) saddle-node separatrix loop, (d) generalized Hopf, (e) neutral saddle separatrix loop, (f) cusp of saddle-nodes of limit cycles. In each of the regions, the limit sets that undergo bifurcations have been sketched by means of diagrams. In (a), one of the nodes on the inner region has been lined through, to help distinguish them and emphasize that the saddle-node curves that meet in the cusp involve different nodes; idem for the limit cycles in (f). In each case, reversing the sign of all the arrows (i.e. inverting all the repulsions) also gives a possible scenario.}
	\label{fig:codimension2}
\end{figure}

This notation is well suited for representing codimension 2 bifurcations as junctions of several codimension 1 curves. The possible codimension 2 bifurcations are shown in Fig.~\ref{fig:codimension2} and can occur generically in two-dimensional parameter spaces \cite{Kuznetsov2004}.

A simple, yet powerful constraint we can apply to the set of bifurcation curves of a system is that whenever several such curves meet at one point in parameter space, all their dressings must match. Every symbol that comes ``in'' through one of the curves must go ``out'' through another, and on the same side of the bifurcation line. Otherwise, a closed path in parameter space that went around the junction of the curves would result in a net creation or destruction of limit sets at each revolution, and the amount and type of limit sets of the system would not be uniquely specified by the parameters.

\section{Examples}\label{sec:examples}

In this section we illustrate how the tools we introduced can be used to help guide a reconstruction of bifurcation diagrams and phase portraits. 

As an example suppose that, from simulations or experiments, a system is known to have a low-dimensional behavior and that, at three different sets of parameters 1, 2 and 3, it has been observed to be respectively stationary, or to oscillate, or to have a coexistence of these two attractors. Moreover, large perturbations of the system tend to die away so that the long term dynamics occurs in a bounded region of phase space. The aim is to suggest plausible bifurcation diagrams in a two-parameter space, and describe the qualitatively different phase portraits that the system would present.

\begin{figure}
	\centering
	\includegraphics[width=.5\linewidth]{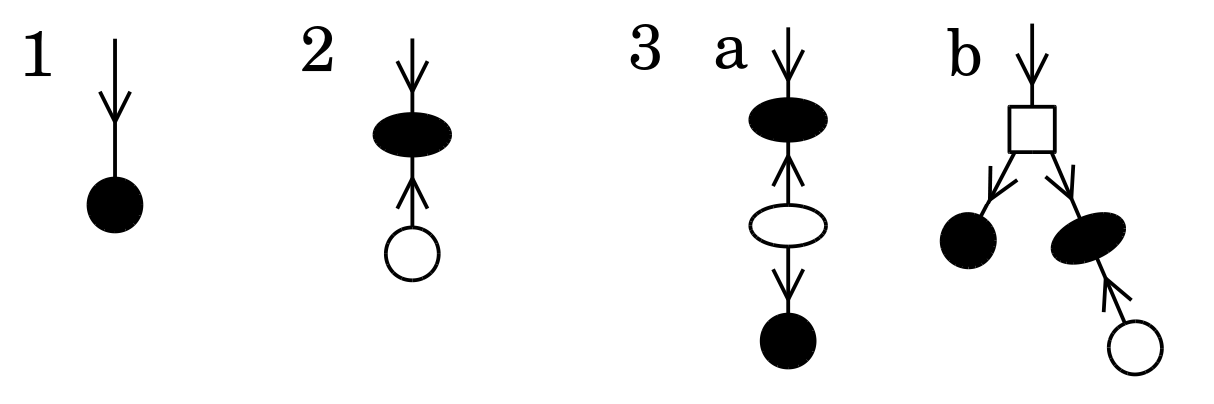}
	\caption{Simplest diagrams featuring (1) a single stable node, (2) a single stable limit cycle, (3) a stable node and a stable limit cycle. Two scenarios are shown in the third case, that lead to different bifurcation diagrams.}
	\label{fig:example}
\end{figure}

The first step is to make complete phase portraits compatible with these sets of attractors. For the first set of parameters, it should have a stable node, for the second, a stable limit cycle, and for the third both, so the diagrams must contain these symbols and eventually other unstable limit sets. Since large perturbations decay, the phase space can be bounded by a closed transversal with repulsion $-1$, and the diagrams should have a downward arrow on top. A single stable node is already a well-formed diagram (since it has exactly one unconnected arrow, at the top) and is the simplest choice for the first case. For the second, at least an unstable node must be added below the limit cycle symbol. For the third case, two relatively simple yet different diagrams can be proposed: by adding an unstable cycle, or a saddle point and an unstable node, as shown in Fig.~\ref{fig:example}. More complex diagrams could be proposed for each case by adding more unstable limit sets. Note, however, that the only way to add only unstable limit sets to a given diagram keeping its index and repulsion constant is by adding pairs of repulsion $-1$ saddle points and unstable nodes.

Let us study the first scenario and suppose that the system has an unstable cycle at the third set of parameters (diagram 3a in Fig.~\ref{fig:example}). There will be regions of parameter space, thus, in which diagrams 1, 2 or 3a apply. This case is simple enough that every diagram can be reached from any of the other two by crossing a single codimension 1 bifurcation, i.e. shrinking a connecting arrow to zero-length. Diagrams 1 and 2 could be bridged with a supercritical Hopf bifurcation, 2 and 3a with a subcritical Hopf, and 3a and 1 with a saddle-node of limit cycles.

One possibility, then, would be that every region is connected to the other two in parameter space, separated by those codimension 1 bifurcations. Indeed, the three bifurcation curves can meet in a generalized Hopf codimension 2 bifurcation, and the bifurcation diagram would look like the one seen in Fig.~\ref{fig:codimension2}d. Or one of the three regions could lie between the other two, then only two codimension 1 curves would be present with no codimension 2 bifurcations.

From the bifurcation diagram it is possible to predict distinct features of the dynamics of the system. From the diagram in Fig.~\ref{fig:codimension2}d, for instance, the period of the limit cycle is expected to be finite everywhere in parameter space, and its amplitude should become arbitrarily small near the Hopf bifurcation. There, a smooth transition should take place between the oscillating and the stationary states. These predictions can be checked against the observed behavior of the system and used to rule out some of the suggested scenarios.

Let us consider the second possibility, in which the third set of parameters corresponds to a diagram like 3b in Fig.~\ref{fig:example}. Now, diagrams 1 and 2 can be connected through a supercritical Hopf as before and 2 and 3b through a saddle-node, but in going from diagram 1 to 3b a stable cycle, an unstable node and a repulsion $1$ saddle should be created, which none of the codimension 1 bifurcations in Table~3 can do. Thus, these regions must be separated by at least two bifurcation curves. An efficient method to find the possible sequences of bifurcations that link two phase portraits in the general case is outlined in the Appendix. The generic bifurcation diagram that one obtains for this scenario is shown in Fig.~\ref{fig:f_example3}. This bifurcation diagram already allows the system to present rich behaviors when the parameters are forced, and it is ubiquitous in numerous dynamical systems: classic examples include the van der Pol \cite{Guckenheimer1983} and the Wilson-Cowan oscillators \cite{Wilson1972}.

\begin{figure}
    \centering
    \includegraphics[width=.5\linewidth]{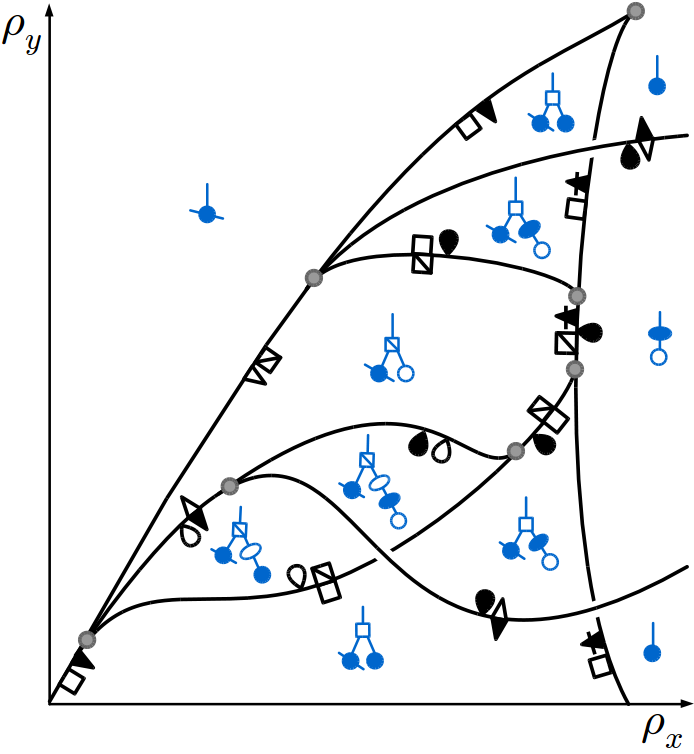}
    \caption{Qualitative bifurcation diagram of the Wilson-Cowan model. The arrows in the diagrams have been omitted for compactness.}
    \label{fig:example5}
\end{figure}

For example, the Wilson-Cowan oscillator describes the mean activities of two coupled populations of neurons, one excitatory and the other inhibitory. A simplified version is given by \cite{Hoppensteadt1997}:
\begin{equation}
	\label{eq:WC}
	\begin{split}
		\dot x &= -x + S(\rho_x + a x - b y) \\
		\dot y &= -y + S(\rho_y + c x - d y),
	\end{split}
\end{equation}
\noindent where $x, y$ are the mean activities of the excitatory and inhibitory populations respectively, $a, b, c, d$ are the couplings between both populations and $\rho_x, \rho_y$ are the external inputs. $S$ is a sigmoidal function, e.g. $S(\xi)=(1+e^{-\xi})^{-1}$. For instance, if we set $a=15$, $b=15$, $c=12$, $d=5$, the resulting bifurcation diagram for the parameters $\rho_x, \rho_y\in(-13,-1)$ has all possible bifurcations of planar systems, as sketched in Fig.~\ref{fig:example5}. This system can have two qualitatively different stationary states, with both populations active or both inactive. The node symbol corresponding to the latter has been lined through in the diagrams of Fig.~\ref{fig:example5} to help distinguish them.

\section{Conclusions}\label{sec:conclusions}
In this work we have treated the problem of constructing phase portraits and bifurcation diagrams of two-dimensional nonlinear systems with a diagrammatic approach.

We introduced a class of diagrams to represent the qualitative features of phase portraits of structurally stable, globally attracting (or repelling) two-dimensional dynamical systems. The diagrams emphasize the robust, topological characteristics of the limit sets of the system, and explicitly discard its quantitative, parameter dependent features. There is a one-to-one correspondence between well-formed diagrams and sets of equivalent (up to topological equivalence and heteroclinic bifurcations) phase portraits. Any phase portrait can be obtained from the diagram and vice versa by following simple algorithmic steps.

Smooth transitions between diagrams give rise naturally to all codimension 1 bifurcations of planar systems (with the exception of heteroclinic connections, which are ignored in our description). We introduced the notion of repulsion of a limit set, an additive quantity that is conserved in all bifurcations, and can be easily computed from a diagram by counting incoming and outgoing arrows. Similarly, the index is also additive, conserved, and can be computed by counting arrows above and below the symbols in a diagram. The global values that these quantities take ($1$ for the index, $\pm1$ for the repulsion) a priori constrain the possible combinations of limit sets that a system can have. These constraints are embedded naturally in the rules for forming diagrams.

We also developed the representation of codimension 1 bifurcations in a two-dimensional parameter space, by adding dressing symbols to the curves. Apart from describing the type of bifurcation, the dressing makes explicit the orientation of a bifurcation curve, i.e. to which side of the curve are the involved limit sets destroyed or created. The dressings are particularly useful for studying codimension 2 bifurcations, which can only occur at a meeting of codimension 1 curves if their dressings match properly.

\section*{Acknowledgements}

This work was supported by CONICET, ANCyT, UBA, and NIH through R01-DC-012859 and R01-DC-006876.

\appendix{Generating sequences of bifurcations}\label{app:fd}

We propose to find sequences of bifurcations that lead from a given phase portrait to another, in a way that is similar to finding contributions to the transition amplitude of a scattering process in Quantum Field Theory (QFT). Our elements will be the limit sets (the ``particles'' in QFT), the codimension 1 bifurcations  will be the ``interaction vertices'' and the sequences of bifurcations the ``Feynman diagrams'' \cite{Feynman1949}. For this reason we will refer to our diagrammatic representation of bifurcation sequences as ``Feynman-like diagrams'' (FL). As with Feynman diagrams in QFT, our representation does not pretend to advance our knowledge of bifurcation theory, but to provide algorithmic tools for keeping track of the possible bifurcation sequences compatible with finite information about the behavior of the system.

The idea is as follows. 
We can represent the limit sets of planar systems with different types of directed lines, as shown in Table~A.1. The direction of the arrow indicates the sign of the repulsion, positive by a left arrow and negative by a right arrow.

\begin{table}[h]
    \centering
    \tbl{Representation by directed lines of the limit sets possible in two-dimensional dynamical systems.}
    {\begin{tabular}{lc}
    	\toprule
    	Limit set & Directed line\\\hline
    	Stable node & 
    	\begin{minipage}{.125\linewidth} % Change eg to .25 for two-column
    		\includegraphics[width=\linewidth]{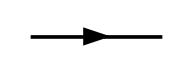}
    	\end{minipage}\\
    	Unstable node & 
    	\begin{minipage}{.125\linewidth}
    		\includegraphics[width=\linewidth]{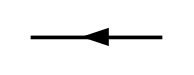}
    	\end{minipage}\\
    	Stable cycle & 
    	\begin{minipage}{.125\linewidth}
    		\includegraphics[width=\linewidth]{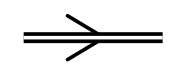}
    	\end{minipage}\\
    	Unstable cycle & 
    	\begin{minipage}{.125\linewidth}
    		\includegraphics[width=\linewidth]{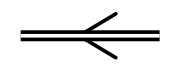}
    	\end{minipage}\\
    	Repulsion $-1$ saddle & 
    	\begin{minipage}{.125\linewidth}
    		\includegraphics[width=\linewidth]{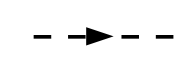}
    	\end{minipage}\\
    	Repulsion $+1$ saddle & 
    	\begin{minipage}{.125\linewidth}
    		\includegraphics[width=\linewidth]{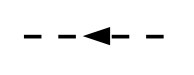}
    	\end{minipage}\\
    	\botrule
    \end{tabular}}
\end{table}

\begin{figure}
	\centering
	\includegraphics[width=.5\linewidth]{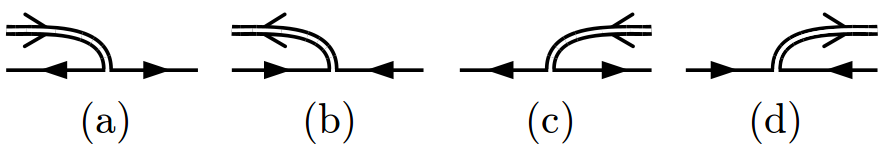}
	\caption{Given a vertex like (a), three more can be obtained by replacing $t\mapsto-t$ (b), $\mu\mapsto-\mu$ (c), or both (d).}
	\label{fig:f_Hopf}
\end{figure}

The limits sets can meet in codimension 1 bifurcations, that are represented by vertices, i.e. particular junctions of lines. These are shown in Table~A.2. There, the horizontal axis represents the bifurcation parameter $\mu$, and the vertical axis the spatial distribution of the limit sets. The interpretation is straightforward: the set of lines is different at both sides of the vertex, which reflects the change in the limit sets of the system at the bifurcation. 
Note that given an allowed vertex, one can obtain three more by reversing the sign of time (i.e. changing the repulsions, or the direction of the arrows), reversing the sign of the bifurcation parameter (interchanging left and right in the vertex), or doing both operations simultaneously. For example, the four possible Hopf bifurcations (supercritical or subcritical, each of which can be crossed in either direction) are explicitly shown in Figure~\ref{fig:f_Hopf}. Accepting that these are allowed operations, we can condense all four variants in a single vertex, as in Table~A.2. Note that, unlike Feynman diagrams, single particle lines cannot be moved from one side of the vertex to the other. In the example of Fig.~\ref{fig:f_Hopf}, an incoming limit cycle could not yield two outgoing nodes due to the conservation of index. Similar considerations apply to the other bifurcations.

\begin{table}
	\centering
	\tbl{The ``vertices'' of two-dimensional dynamics. Each represents a codimension 1 bifurcation. The allowed variants are illustrated in Fig.~\ref{fig:f_Hopf}.}
	{\begin{tabular}{lc}
		\toprule
		Bifurcation & Vertex \\\hline
		Saddle-Node & 
		\begin{minipage}{.17\linewidth}\centering % Use eg .27 in two-column
			\includegraphics[width=\linewidth]{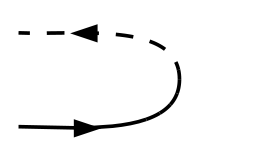}
		\end{minipage}\\
		Hopf & 
		\begin{minipage}{.17\linewidth}\centering
			\includegraphics[width=\linewidth]{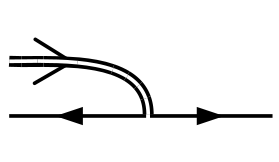}
		\end{minipage}\\
		\begin{tabular}{@{}l@{}}Saddle-Node on\\Invariant Cycle\end{tabular} & 
		\begin{minipage}{.17\linewidth}\centering
			\includegraphics[width=\linewidth]{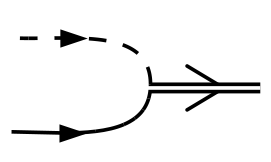}
		\end{minipage}\\
		Homoclinic &
		\begin{minipage}{.17\linewidth}\centering
			\includegraphics[width=\linewidth]{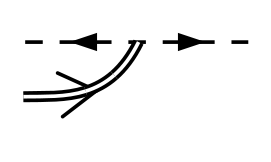}
		\end{minipage}\\
		\begin{tabular}{@{}l@{}}Saddle-Node of\\Limit Cycles\end{tabular} & 
		\begin{minipage}{.17\linewidth}\centering
			\includegraphics[width=\linewidth]{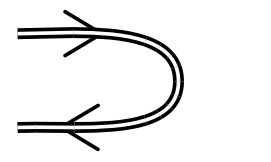}
		\end{minipage}\\
		\botrule
	\end{tabular}}
\end{table}

A sequence of bifurcations with parameter $\mu$ is represented by a FL diagram containing several vertices. The lines present at each value of $\mu$ give the succession of limit sets at each stage. The possible sequences of bifurcations between two phase portraits are given by all FL diagrams whose external lines correspond to the initial and final portraits. Provided that these have the same total repulsion, there will be an infinite number of possible sequences that link them. As in QFT, it is reasonable to classify them by the number of bifurcations (vertices) they involve. The motivation for that is that sequences with few bifurcations offer the most parsimonious scenarios and are more limited in number. In perturbative QFT, diagrams with fewer vertices correspond to lower orders in perturbation theory and usually contribute more to the scattering amplitude.

\begin{figure}
    \centering
    \includegraphics[width=.35\linewidth]{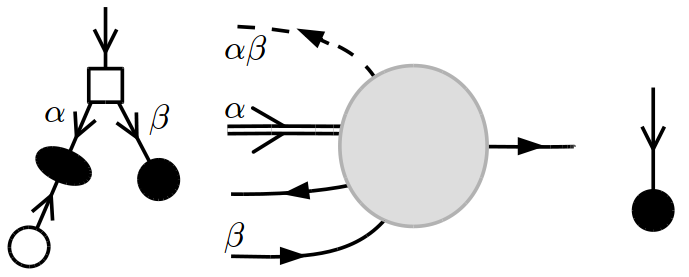}
    \caption{A path in parameter space between two qualitatively different regions defines a series of bifurcations, represented by a FL diagram. Here the blob represents any diagram featuring the depicted external lines. The incoming lines encode the limit cycles of the initial region, and the outgoing lines, of the final region.}
    \label{fig:f_example1}
\end{figure}

We will illustrate the procedure with the transition between regions 1 and 3b in the example considered in Fig.~\ref{fig:example}. The first step is to identify the initial and final limit sets, which will be the external lines of the FL diagrams. In a diagram representing a transition from region 3 to 1, these would look as shown in Fig.~\ref{fig:f_example1}. The rationale we take for the vertical ordering of multiple lines is the same we used for the diagrams introduced in Section~\ref{ssec:construction}. As we have seen,  in these diagrams saddle points emanate ``branches'' downwards. To keep track of their connectivities, we can label them with dummy symbols $\alpha$, $\beta$, $\ldots$ and use the labels as reference marks, as on the left side of Fig.~\ref{fig:f_example1}. There, branches $\alpha$ and $\beta$ stem from the saddle point, branch $\alpha$ has a limit cycle on top and branch $\beta$ only has a node. With this ordering convention, bifurcations can only occur between neighboring lines, including those linked by the reference marks. The case of multiple saddle points is incorporated by allowing adjacent saddle-point lines to interchange their positions and/or repulsions. 

\begin{figure}
	\centering
	\includegraphics[width=.45\linewidth]{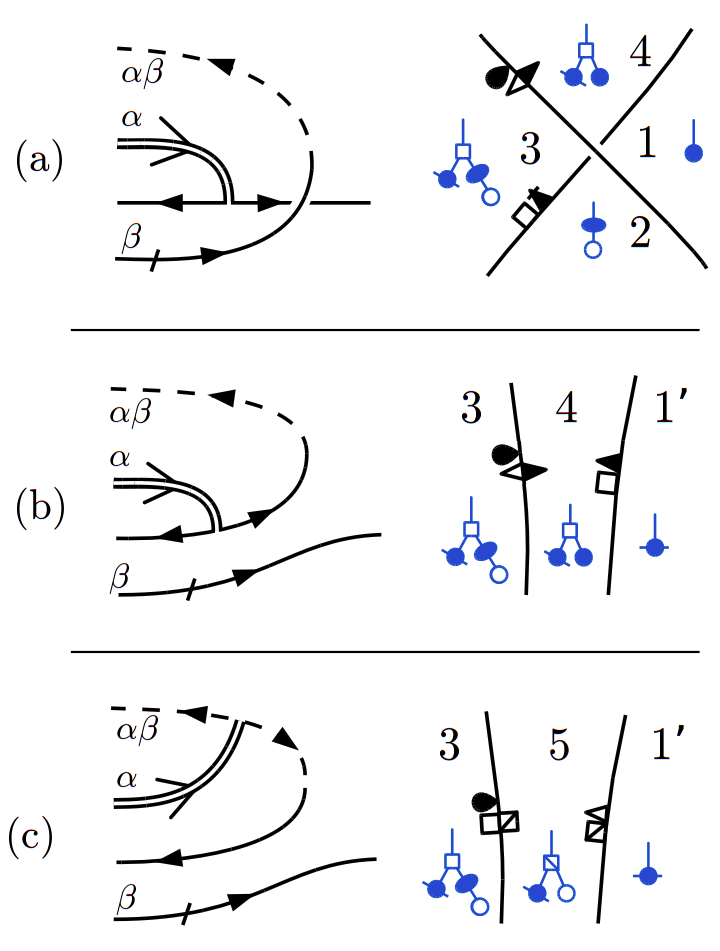}
	\caption{\textit{Left panels:} the three possible FL diagrams linking regions 3 and 1 with the minimal number of vertices (two). \textit{Right panels:} sketch of how the bifurcation curves associated with these diagrams would look in a two-parameter space. They define regions, numbered 1-5, in which the phase portraits are given by the diagrams drawn in blue. The node in branch $\beta$ has been lined through to help distinguish it from the other.}
	\label{fig:f_example2}
\end{figure}

The next step is to find, in an orderly manner, well-formed FL diagrams featuring these external lines. As argued above, it is reasonable to look at the diagrams with smaller number of vertices first. In this particular case, the initial and final lines cannot be linked by a single bifurcation, a minimum of two must be used. There are three different possible diagrams with two vertices, that are shown in Fig.~\ref{fig:f_example2}.

Note that in case (a) in Fig.~\ref{fig:f_example2}, the saddle-node and Hopf bifurcations involve different limit sets, so they may occur in either order. Thus, in the general case these bifurcation curves could intersect in parameter space, as shown in the right panel. Diagram (a) can be interpreted as a path in parameter space that goes from region 3 to region 1 crossing these bifurcation curves. If the saddle-node occurs first, it would traverse an intermediate region with an unstable node surrounded by a stable cycle, which we can recognize as the region 2 introduced in Fig.~\ref{fig:example}. If the Hopf occurs first, a new region appears, labeled 4. There, two stable fixed points coexist, which is a testable prediction of this bifurcation diagram. A typical example of a system exhibiting such coexistence is an ``on-off'' switch. Note that the intersection of these two curves represents two independent codimension 1 bifurcations rather than a bifurcation of intrinsic codimension 2.

On the contrary, in (b) both bifurcations involve the same node, and the Hopf bifurcation must occur before the saddle-node. The path in parameter space would traverse region 4 again but, unlike case (a), the node that survives is now the one on branch $\beta$. Since these nodes represent different states of the system (e.g. ``on'' or ``off''), we label the final region $1'$ to distinguish it from the former. Notice, however, that this is a merely quantitative difference and that it could be possible to transform one state into the other continuously. 

Similarly, in (c) both bifurcations involve the same saddle point and the homoclinic must occur first. The final region is again $1'$ since the surviving node is the one from branch $\beta$.

\begin{figure}
	\centering
	\includegraphics[width=.4\linewidth]{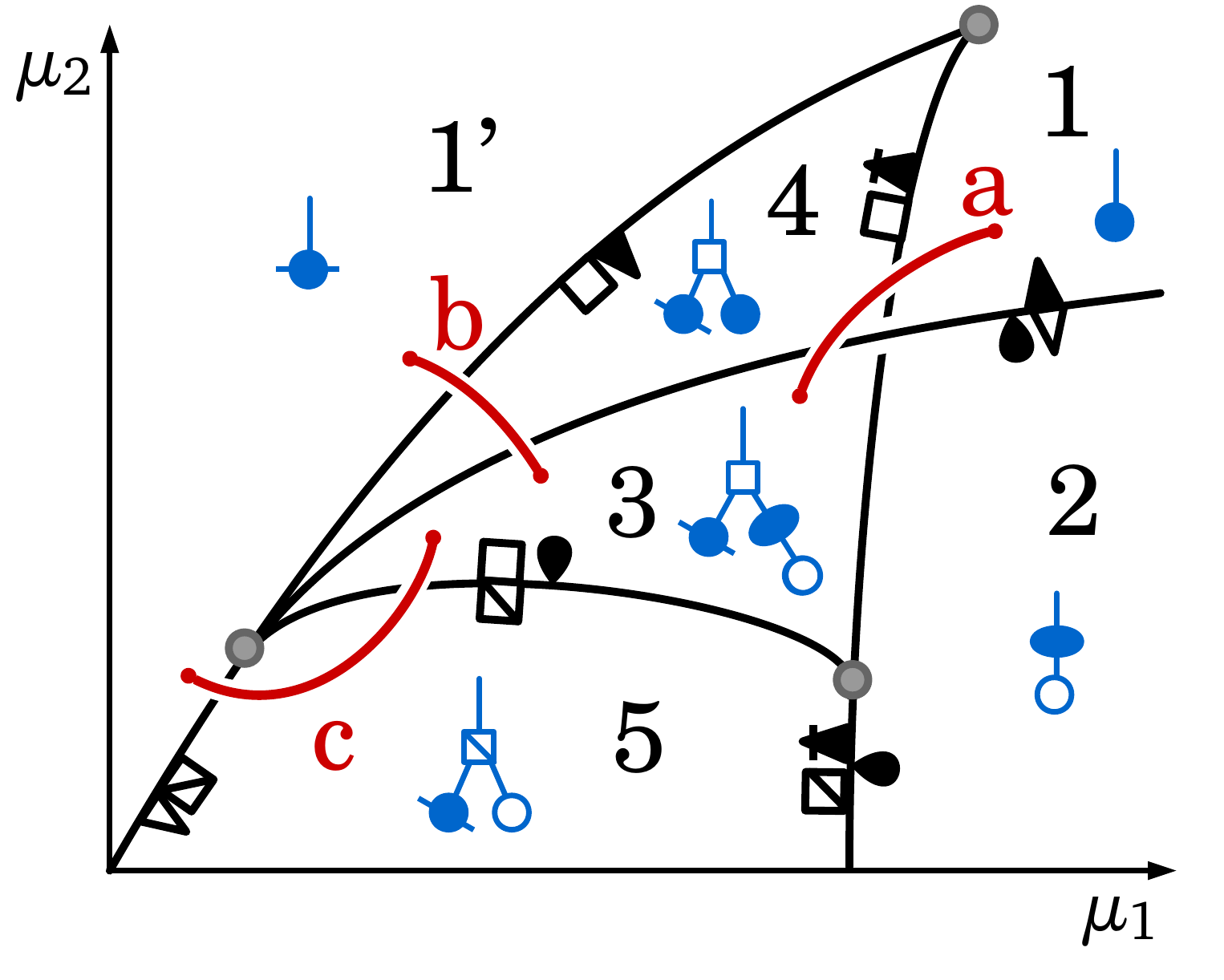}
	\caption{
	Bifurcation diagram for the second scenario in Fig.~\ref{fig:example}, describing the possible transitions between regions 1 and 3 that involve series of two codimension 1 bifurcations. It features a cusp, Bogdanov-Takens and saddle-node separatrix loop codimension 2 bifurcations (emphasized with circles). The labeling of the regions 1-5 and paths a-c is consistent with Fig.~\ref{fig:f_example2}.}
	\label{fig:f_example3}
\end{figure}

The final step is to combine the information from the three diagrams of Fig.~\ref{fig:f_example2} in a single bifurcation diagram. We see that region 3 should be adjacent to regions 2, 4 and 5. In turn, regions 2 and 4 limit with 1, and regions 4 and 5 with $1'$. The resulting bifurcation diagram is shown in Fig.~\ref{fig:f_example3}, where we have obtained the junctions of curves (emphasized by gray circles) from the codimension 2 bifurcations of Fig.~\ref{fig:codimension2}. Note that the dressings of the colliding bifurcation curves matched properly in every case. If the homoclinic born at the Bogdanov-Takens bifurcation reaches the saddle-node curve separating regions 2 an 3, it must collide with it at a saddle-node separatrix loop bifurcation. It cannot cross it, since on the other side lies region 2, that has no saddle point to undergo an homoclinic bifurcation. And indeed, regions 2 and 5 can be connected by the resulting saddle node in invariant cycle bifurcation. The three paths a, b and c correspond to the three sequences of bifurcations encountered in Fig.~\ref{fig:f_example2}. This bifurcation diagram describes general transitions between regions 1 and 3 with up to two codimension 1 bifurcations, in the sense that any other such bifurcation diagram will typically be possible to map to a part of this one. We note that not all bifurcations shown here must necessarily be accessible to a specific system (for example, the two saddle-node curves need not meet in the cusp, they could extend further). More complex sequences of bifurcations can be taken into account in a similar manner by drawing FL diagrams with increasing number of vertices.

\bibliographystyle{ws-ijbc}
\bibliography{references}

\end{document}